\documentclass{article}
\usepackage{spconf,amsmath,graphicx,setspace,enumitem}
\usepackage{amssymb}
\usepackage{bm}
\usepackage{subcaption, url}
\usepackage[dvipsnames]{xcolor}
\usepackage{tikz}
\usepackage{lipsum}
\usepackage[absolute]{textpos}

\newcommand\copyrighttext{%
  \footnotesize © 2022 IEEE. Personal use of this material is permitted. Permission from IEEE must be
obtained for all other uses, in any current or future media, including
reprinting/republishing this material for advertising or promotional purposes, creating new
collective works, for resale or redistribution to servers or lists, or reuse of any copyrighted
component of this work in other works.}
\newcommand\copyrightnote{%
  \begin{tikzpicture}[remember picture,overlay]
    \node[anchor=south,yshift=10pt,draw] 
    at (current page.south) 
    {\parbox{\dimexpr\textwidth-\fboxsep-\fboxrule\relax}{\copyrighttext}};
  \end{tikzpicture}%
}

\title{Upmixing via style transfer: a variational autoencoder for disentangling spatial images and musical content}
\name{Haici Yang$^1$\thanks{ The paper describes work performed at Amazon.}, Sanna Wager$^{2}$, Spencer Russell$^{2}$, Mike Luo$^2$, Minje Kim$^{1,2}$, Wontak Kim$^2$}
\address{ $^1$Indiana University, Dept. of Intelligent Systems Engineering, Bloomington, IN, USA\\
  $^2$Amazon Lab126, Cambridge, MA and Sunnyvale, CA, USA}

\begin{document}

\ninept
\maketitle
%
\begin{abstract}
In the stereo-to-multichannel upmixing problem for music, one of the main tasks is to set the directionality of the instrument sources in the multichannel rendering results. In this paper, we propose a modified variational autoencoder model that learns a latent space to describe the spatial images in multichannel music. We seek to disentangle the spatial images and music content, so the learned latent variables are invariant to the music. At test time, we use the latent variables to control the panning of sources. We propose two upmixing use cases: transferring the spatial images from one song to another and blind panning based on the generative model. 
We report objective and subjective evaluation results to empirically show that our model captures spatial images separately from music content and achieves transfer-based interactive panning.
\end{abstract}

\copyrightnote

\section{Introduction}
\label{sec:intro}
While home theater and surround sound have become more common and affordable, most music is still in stereo. Approaches to play stereo content on multichannel speakers can be categorized into two kinds. The first uses matrix \cite{gerzon1991optimal} to map the left and right content channels into each of the surround speakers. These operate in different sound modes, e.g. direct, stereo, and all channel stereo mode, without sources decomposition. The second approach relies on the separation of sound components from the stereo mixture, e.g., into music content and other ambient sounds \cite{6082279, faller2006multiple} or into individual musical instruments \cite{jeon2010robust} that are then placed in the desired panning locations. There are different techniques proposed to the primary-ambient separation, including frequency-domain inter-channel coherence index computation \cite{5745013,avendano2002frequency,merimaa2007correlation-based}, time or frequency-domain adaptive-filtering \cite{irwan2002two, usher2007enhancement}, PCA-based primary-ambient decomposition \cite{baek2012efficient, ibrahim2016primary}, and deep neural network (DNN)-based methods \cite{uhle2008supervised, ibrahim2018primary, choi2021exploiting}. Perceived azimuth direction has also proven to be beneficial for direct and ambient separation \cite{kraft2015stereo}. Jeon \textit{et al}. proposed an upmixing algorithm based on robust source separation with a post-scaling algorithm to compensate for the interference \cite{jeon2010robust}. Another DNN-based upmixing model, introduced by Park \textit{et.al} \cite{park2016subband}, bypasses any explicit decomposition and generates the center and surround channels from stereo channels directly based on non-linear transformations learned in the DNN. 


In this paper, we develop a generative model for  upmixing. The goal is to implicitly separate the instrumental components contained in the stereo signal and rearrange them into the five-channel setting that renders the desired perceptual source directions. This setting of the problem leads to the fact that there can be many legitimate upmixed versions that correspond to a stereo signal, thus making optimization ill-defined on this one-to-many mapping function . 

To address this issue, we firstly make a basic assumption, that the spatial images of a multi-channel music signal can be represented independently of the musical content. This means that two different 5-channel music signals with the same instrumentation and panning will map to an identical spatial representation even though the content—melody, harmony, etc—differs. Then, we hope to extract the source-specific spatial representation from the multichannel audio and utilize the learned representation to guide the test-time upmixing.
To this end, we propose a modified variational autoencoder (VAE) model \cite{KingmaD2014vae} and train it in a supervised manner. Similar to the ordinary VAE training, we use a 5-channel signal as both input and target. However, we also provide a stereo version of the 5-channel signal as input to the decoder and train the VAE bottleneck layer to exclusively capture the spatial images of the sources rather than an entangled representation of music components and spatial information.

We empirically show that the learned latent representation reflects the spatial images of the multichannel input, i.e., it is correlated to the panning strategy used to render the 5-channel input, and is invariant to the music content. Hence, the decoder of the trained VAE can generate the upmixed version of the input stereo with the guidance provided in the form of the latent variables. Moreover, we also show that panning information represented in the latent space can be transferred in-between different songs, enabling spatial style transfer.



\section{Upmixing: a probabilistic formulation}


We firstly describe a general probabilistic upmixing model where spatial information and the stereo music signal are still entangled. Then, we showcase how the assumption that spatial information and stereo audio are independent alters the proposed formulation. We further introduce an alternative deterministic downmixing process, to achieve the desired disentanglement.

We define the upmixing process probabilistically, as the likelihood of observing the 5-channel sample $\bar{\bm{X}}^{(5)}$ given the spatial information $\bar{\bm{h}}$ and stereo audio $\bar{\bm{X}}^{(2)}$: $p_\theta(\bm{X}^{(5)}|\bm{X}^{(2)},\bm{h})$, where $\theta$ stands for the model parameters and the bar notation is the sample of a random variable, e.g., $\bar{\bm{X}}^{(5)}\sim\bm{X}^{(5)}$. We use superscript to denote the number of channels. 
However, we cannot compute the likelihood due to the intractable marginal distribution $p(\bm{X}^{(5)})$ and posterior distribution $p(\bm{X}^{(2)}, \bm{h}|\bm{X}^{(5)})$. VAE employs variational inference to resolve this issue by proposing an encoder function $q_\phi(\bm{X}^{(2)}, \bm{h}|\bm{X}^{(5)})$ that best approximates the true posterior. Coupled with concept of evidence lower bound (ELBO), the optimization is eventually defined as follows:
\begin{equation}
 \begin{split}\label{first}
    \mathcal{L}(\theta, \phi; \bm{X}^{(5)}) 
    =  &  \mathbb{E}_{q_\phi(\bm{X}^{(2)}, \bm{h}|\bm{X}^{(5)})}[\log p_\theta(\bm{X}^{(5)}|\bm{X}^{(2)}, \bm{h})]\\
    - & D_{KL}\big(q_\phi(\bm{X}^{(2)}, \bm{h}|\bm{X}^{(5)}) ||
    p(\bm{X}^{(2)}, \bm{h})\big),
 \end{split}
\end{equation}
where $p(\bm{X}^{(2)}, \bm{h})$ denotes the prior distribution of the latent variables, thus forming a regularization term. Meanwhile, the main reconstruction objective is to maximize the expectation of the log-likelihood w.r.t. the approximated posterior distribution. 

The desired disentanglement of stereo music content and spatial information assumes that $\bm{X}^{(2)}$ and $\bm{h}$ are independent from each other. Therefore, the joint posterior distribution can be factorized, i.e., $p(\bm{X}^{(2)}, \bm{h}|\bm{X}^{(5)}) = p(\bm{X}^{(2)}|\bm{X}^{(5)})p(\bm{h}|\bm{X}^{(5)})$. To further constrain that $\bm{X}^{(2)}$ only reflects stereo music content while $\bm{h}$ represents 5-channel spatial information, we employ a hard regularization trick by replacing the random variable $\bm{X}^{(2)}$  with a simple downmixed version $\bar{\bm{X}}^{(2)}$. Then, the probabilistic model $p(\bm{X}^{(2)}|\bm{X}^{(5)})$ is simplified into a deterministic downmixing process, $\bar{\bm{X}}^{(2)}=\mathcal{G}(\bar{\bm{X}}^{(5)})$, where $\mathcal{G}$ denotes the pre-defined downmix function. We treat $p(\bm{X}^{(2)}|\bm{X}^{(5)})$ as a constant and rewrite eq. \eqref{first} as follows: 
 \begin{equation}\label{propto}
 \begin{split}
    \mathcal{L}(\theta, \phi; \bm{X}^{(5)}, \bm{X}^{(2)})
    \propto  &\mathbb{E}_{q_\phi(\bm{h}|\bm{X}^{(5)})}[\log p_\theta(\bm{X}^{(5)}|\bar{\bm{X}}^{(2)}, \bm{h})] \\
    - &D_{KL}\big(q_\phi(\bm{h}|\bm{X}^{(5)})||p(\bm{h})\big).  
    \end{split}
 \end{equation}

The loss function represents our proposed disentanglement method. As a modified VAE, the spatial information is learned via $\bm{h}$ while the decoder takes in extra stereo input to generate the multi-channel output $\bm{X}^{(5)}$.
The second term regularizes each latent variable by a standard normal distribution, i.e., $p(\bm{h}) = \mathcal{N}(\bm{0},\bm{I})$, where $\bm{I}$ is a $\!J$ dimensional identical matrix.


\begin{figure}[t]
\begin{center}
	\includegraphics[width=\columnwidth]{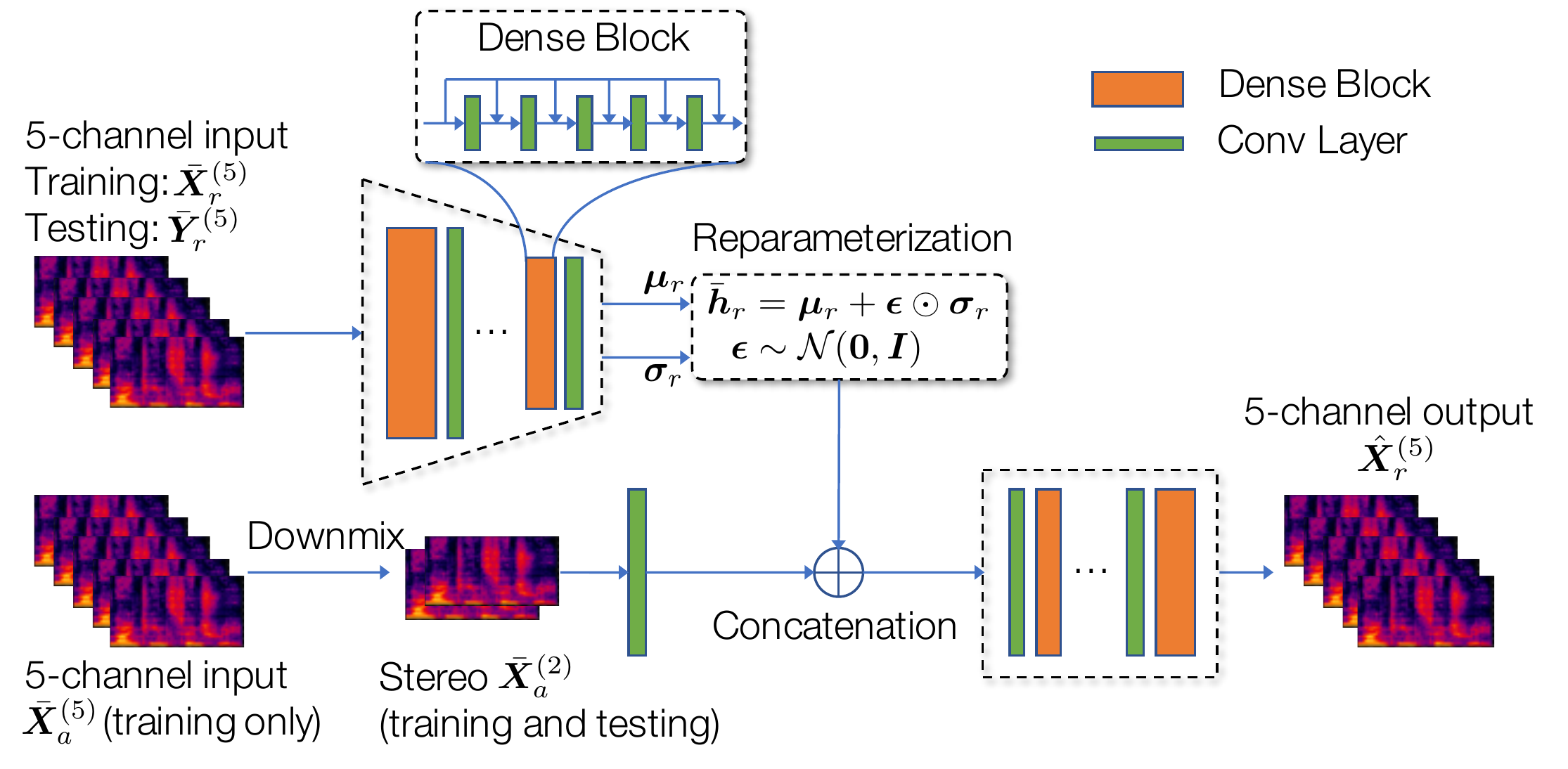}
\caption{Model architecture}
\label{architecture}
\end{center}
\end{figure}

\section{Model Description}
\subsection{DNN architecture}
Fig. \ref{architecture} summarizes the overall model architecture. 
We employ densely connected convolutional blocks introduced in DenseNet \cite{HuangG2017densenet} as our building blocks in both the encoder and the decoder.
In both modules, we alternate 5 dense blocks with 4 transition layers. A dense block contains 5 convolutional layers, each of which produces $C=20$ output channels. We set the stride to be 2 for the transition layers in the encoder to reduce the data dimension. During training, the encoder takes as input a 5-channel audio in the form of stacked magnitude spectrograms, $\bar{\bm{X}}_r^{(5)}\in\mathbb{R}_+^{5\times F\times T}$, where $F$ and $T$ are the frequency subbands and time-domain frames. Here, we introduce a subscript $r$ to denote the particular panning strategy for creating a 5-channel input signal. The encoder outputs two $J$-dimensional vectors $\bm{\mu}_r, \bm{\sigma}_r\in\mathbb{R}^{J}$, which are the parameters of the $J$-dimensional multivariate latent normal distribution. Latent variables are then sampled using the reparameterization technique \cite{KingmaD2014vae}, i.e., $\bar{\bm{h}}_r=\bm{\mu}_r+\bm{\epsilon}\odot\bm{\sigma}_r$ where $\bm{\epsilon} \sim \mathcal{N}(\bm{0},\bm{I})$.



The stacked stereo spectrogram, $\bar{\bm{X}}^{(2)} \in \mathbb{R}^{2\times F\times T}$, is a unique feature of the proposed model. It is input to the decoder to provide the music signal for upmixing. During training, we first prepare a 5-channel signal $\bar{\bm{X}}_r^{(5)}$ under a panning condition $r$, from which the encoder learns the latent vector ${\bm{h}}_r$. Meanwhile, the same music stems are rearranged into another 5-channel version $\bar{\bm{X}}_a^{(5)}$ using a different panning condition $a$, whose stereo downmix $\bar{\bm{X}}_a^{(2)}$ is fed to the decoder. We rearrange the source locations in $\bar{\bm{X}}_a^{(2)}$ rather than reusing $\bar{\bm{X}}_r^{(2)}$, to ensure that the left- and right panning in the stereo is unrelated to the left- and right panning in the 5-channel signal.



We repeat $\bar{{\bm{h}}}_r$ and stack to $\bar{\bm{X}}_a^{(2)}$ at the concatenation step to condition the spatial information on the stereo input.
The transition layer that receives the stereo $\bar{\bm{X}}_a^{(2)}$ does not change the tensor size, which is $2\!\times\!F\!\times\!T$. After repeating the values of $J$-dimensional embedding vector $\bar{{\bm{h}}}_r$ at all $F\!\times\!T$ bins, the input to the decoder is of size $(2\!+\!J)\!\times\!F\!\times\!T$. 


\subsection{The Test-Time Generation Processes}\label{sec:test}
For test-time generation, the decoder operates as the upmix generator, which takes stereo signal $\bm{X}_a^{(2)}$ and the spatial information ${\bm{h}}_r$ as a seed. Our model can accommodate two use-case scenarios:

\begin{itemize}[noitemsep,topsep=0pt, leftmargin=0in, itemindent=.15in]
\item \textbf{Style transfer-based upmixing}: We posit that panning style transfer from one music piece to another is possible. Here we define the panning style as the set of apparent source directions for each instrument. In this case, the user provides two input signals: $\bar{\bm{Y}}_s^{(5)}$ as the source 5-channel audio with the panning style $s$ and $\bar{\bm{Z}}^{(2)}$ as the stereo music to upmix. Note that during the test time we do not specify the panning method of $\bar{\bm{Z}}^{(2)}$ as it doesn't influence the process. The actual music in  $\bar{\bm{Y}}_s^{(5)}$ and $\bar{\bm{Z}}^{(2)}$ differs and the aim of the generator is to create a 5-channel audio $\hat{\bm{Z}}_s^{(5)}$ which is upmixed from $\bar{\bm{Z}}^{(2)}$ and has the same panning method as $\bar{\bm{Y}}_s^{(5)}$.  The proposed style transfer is conducted by sharing the latent variables $\bm{h}_s$ in style extraction (i.e. encoding) and upmixing (i.e. decoding) processes. The encoder takes in $\bar{\bm{Y}}_s^{(5)}$ and learns $\bm{h}_s$ which encodes the source's spatial panning. The generator takes in stereo seed signal $\bar{\bm{Z}}^{(2)}$ and $\bar{\bm{h}}_s$,  and synthesizes the multi-channel version of the target $\hat{\bm{Z}}_s^{(5)}$.  
This indirect way enables a user interface where the user handles spatial control by providing an example song. 

\item \textbf{Blind upmixing}: The blind upmixing method literally ``generates" spatial panning information from the learned latent variables $\bm{h}$ and applies it to the seed stereo signal $\bar{\bm{Z}}^{(2)}$. In our VAE, the generation process begins with a random sample $\bar{\bm{h}}$ only, hence the spatial image of the generated 5-channel audio sample $\hat{\bm{Z}}^{(5)}$ is out of the user's control. In the blind upmixing scenario, the sampling process is completely random without involving any intuition about the latent space. 
\end{itemize}

\subsection{The Baseline}
Although blind upmixing has been studied extensively in the literature, our approach to upmixing via style transfer is novel. Hence,
we are unable to use any existing model as a baseline. 
Instead, we build a simple baseline model by spreading each channel of $\bar{\bm{Z}}^{(2)}$ to the front and rear channels of the same side in the 5-channel output. This is the same as the ``all channel stereo'' upmixing
mode commonly used by home receivers. It is a straightforward way to perform upmixing and can perfectly preserve the sound quality of the stereo input because the process does not generate any artifacts. By comparing with the baseline model, we mainly seek to prove the functionality of our model on style-transfer upmixing. 

\section{Experimental Setup}
\subsection{Vector-Based Amplitude Panning for Data Preparation}
\label{VBAP}

To train and evaluate the model, it is necessary for us to know the panning method that is used to create the 5-channel audio, so as to examine whether the information captured in the latent space is correlated to the correct spatial images that we want to learn. Also, when having stem sources, we can expose the model to various artificial upmixing configurations where source locations change freely. However, such ground-truth spatial maps are not readily available for most 5-channel music signals we have access to. 

To that end, we choose to build our own 5-channel dataset from individual musical instruments via the vector-based amplitude panning (VBAP) method \cite{pulkki1997virtual}. VBAP provides an efficient equation for virtual sound source positioning, enabling control of an unlimited number of loudspeakers in an arbitrary two- or three-dimensional placement around the listener. In this paper, we employ a two-dimensional rendering space  with 5 speakers, following the ITU's 5.1 channel standard \cite{ITU5.1} without the subwoofer, which is not considered in our upmix algorithm.
For each instrument, we specify a virtual source direction. Then, we pan each instrument independently using the two adjacent speakers to the desired coming direction, based on the vector base formulation \cite{pulkki1997virtual}.


\subsection{Datasets}

We build a synthesized 5-channel dataset using the MUSDB18 dataset \cite{musdb18-hq}. MUSDB18 provides pop songs in stereo format and four separated instrumental stems: \texttt{vocals}, \texttt{drums}, \texttt{bass}, and \texttt{other}. We split data set into training, validation, and testing subsets, which amount to about 5, 0.5, and 1.5 hours, respectively. The 5-channel versions are created using the VBAP method described in Sec. \ref{VBAP} by controlling the stem tracks. Stereo input signals $\bar{\bm{X}}^{(2)}$ are rendered by applying passive downmixing  on the synthesized 5-channel, i.e., $\bar{\bm{X}}^{(2)}=\mathcal{G}(\bar{\bm{X}}^{(5)})$, where $\bar{\bm{X}}^{(2)}_{[1,:,:]} = \bar{\bm{X}}^{(5)}_{[1,:,:]} + \bar{\bm{X}}^{(5)}_{[2,:,:]} + \bar{\bm{X}}^{(5)}_{[3,:,:]}/2$ and $\bar{\bm{X}}^{(2)}_{[2,:,:]} = \bar{\bm{X}}^{(5)}_{[4,:,:]} + \bar{\bm{X}}^{(5)}_{[5,:,:]} + \bar{\bm{X}}^{(5)}_{[3,:,:]}/2$. The order of the five channels in $\bar{\bm{X}}^{(5)}$ are : front left, real left, center, front right, and real right. 

We mainly run our experiments on the synthesized MUSDB18 dataset. However, when validating the blind upmixing model, we employ another internal real-world dataset of professionally mixed surround music with approximately 17 hours for training, 2.5 hours for validation, and 4 hours for testing, in order to test our model's generalization ability on real music data.



\subsection{Training Setup}

The directions of sources are randomly sampled from the entire $360^\circ$ circle. 
The models are trained on 2.2 second-long segments,  which have at least one instrument not silent. Each segment is processed by short-time Fourier transform (STFT) on Hann-windowed frames of 1024 samples with a 75\% overlap, resulting in a spectrogram of size $513\times 384$. We apply the phase of the stereo input to recover 5-channel magnitude spectrograms. The left channel's phase in stereo is used for both front and rear left channels in the 5-channel audio, and similarly for the right channels; the mean of the left and right channels is used for the center channel recovery. 
We use the Adam optimizer with an initial learning rate of 0.005 \cite{KingmaD2015adam}.

\subsection{Objective Evaluation Methods}


We evaluate our style transfer-based upmixing model against the baseline using various objective metrics as follows:
\begin{itemize}[itemsep=1pt,topsep=0pt,leftmargin=*]
\item \textbf{Overall reconstruction quality}: Scale-dependent source-to-distortion ratio (SD-SDR) is employed to report the overall reconstruction quality by comparing the time-domain reconstruction $\hat{\bm{z}}_s^{(5)}$ and the ground-truth ${\bm{z}_s^{(5)}}$ as SD-SDR is proven to better reflect the scale reconstruction compared to ordinary SDR \cite{LeRouxJL2018sisdr}. 
\item \textbf{Spatial reconstruction quality}: To validate the spatial reconstruction quality, we propose a new metric using the Wasserstein distance on interchannel level differences (WILD). Given the 5-channel magnitude spectrogram $\bm{Z}\in\mathbb{R}_+^{5\times F\times T}$, we compute an ILD matrix for each pair of channels $(i,j)$, e.g., $\bm{M}^{(i,j)}_{[f,t]}=20\log_{10}\big(\|\bm{Z}_{[i,f,t]}\|/\bm{\|Z}_{[j,f,t]}\|\big)$. The five channels result in ten such ILD matrices in total, i.e., $\bm{M}\in\mathbb{R}^{10\times F\times T}$. 
The Wasserstein distance computes on the histograms of target and estimation ILD matrices, i.e., $\sum_{k=1}^{10} \text{WILD}\big(\text{hist}(\bm{M}_{[k,:,:]}), \text{hist}(\hat{\bm{M}}_{[k,:,:]})\big)$. Note that the Wasserstein distance improves the robustness of the comparison when the two histograms are too different from each other, where KL-divergence fails to quantify the dissimilarity. 
\item \textbf{Average angle difference}: We use the least square algorithm to approximately decompose each of the 5 channels into a linear combination of four instrumental sources. Based on the relative amount of each instrument spread in the 5 channels, we can estimate the virtual location of the instrument. We report the difference between the ground-truth and estimated direction of the upmixed instruments in terms of angles in degree. 
\end{itemize}

\subsection{Subjective Evaluations}
We conduct an ABX test to subjectively evaluate the proposed spatial style transfer algorithm. The ground-truth 5-channel version $\bm{Z}_s^{(5)}$ is provided to the participants as the reference. Participants are asked to choose the one that is more similar to $\bm{Z}_s^{(5)}$ between the style-transferred reconstruction $\bm{\hat{Z}}_s^{(5)}$ and the baseline upmix $\bm{\tilde{Z}}^{(5)}$, in terms of the  incoming direction of the four sources separately, as well as the overall spatial image. The test contains 5 trials and is conducted in a professional surround-sound listening room.

\section{Experimental Results}
\subsection{Analysis on the Learnt Latent Space}

\begin{figure}[t]
     \centering
     \begin{subfigure}[b]{\columnwidth}
        \begin{subfigure}[b]{0.495\columnwidth}
             \includegraphics[width=\textwidth]{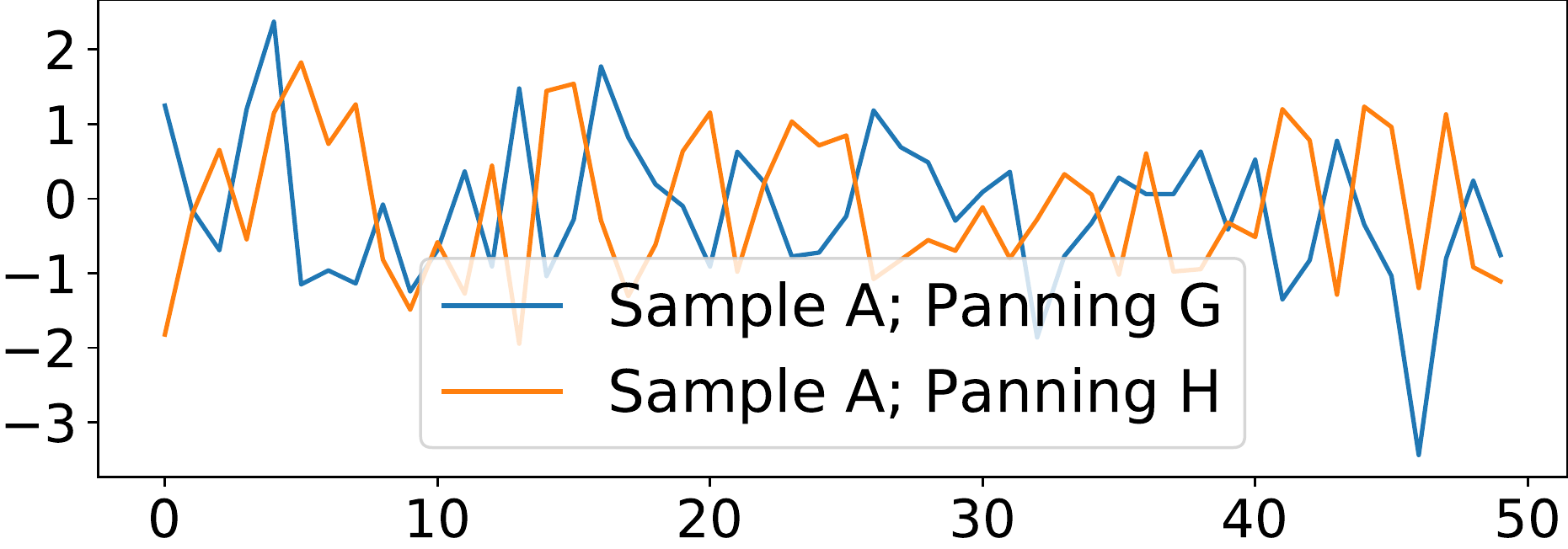}
             \caption{}
             \label{fig:a}
        \end{subfigure}
        \begin{subfigure}[b]{0.495\columnwidth}
             \includegraphics[width=\textwidth]{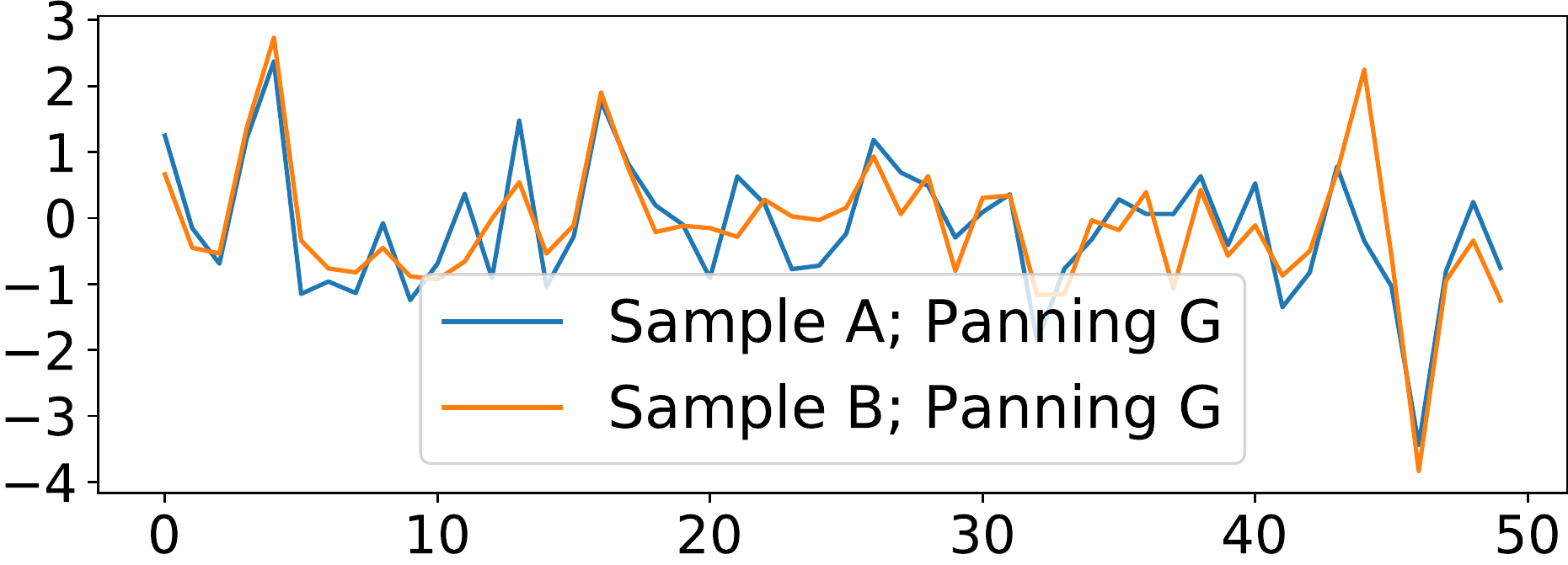}
             \caption{}
             \label{fig:b}
        \end{subfigure}
     \end{subfigure}
     \begin{subfigure}[b]{\columnwidth}
        \begin{subfigure}[b]{0.48\columnwidth}
             \centering
             \includegraphics[width=\textwidth]{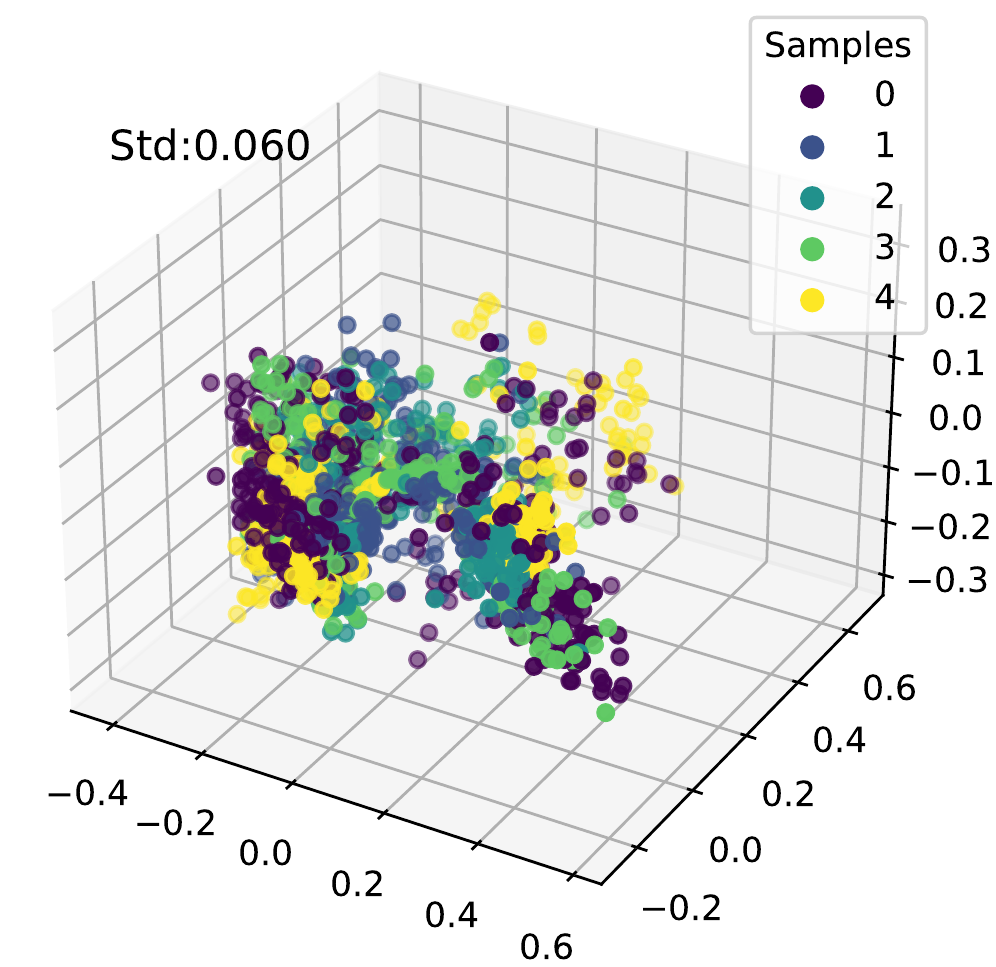}
             \caption{}
             \label{fig:c}
        \end{subfigure}
        \begin{subfigure}[b]{0.48\columnwidth}
             \centering
             \includegraphics[width=\textwidth]{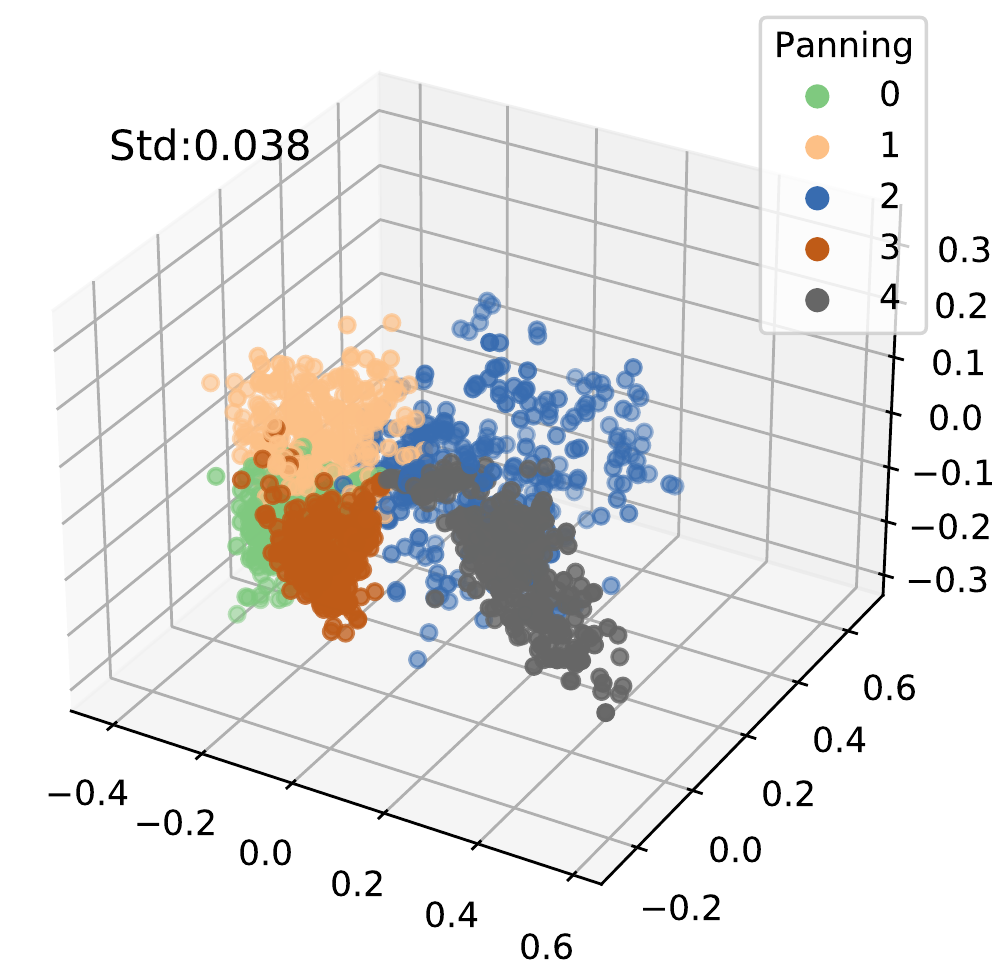}
             \caption{}
             \label{fig:d}
        \end{subfigure}
     \end{subfigure}
    \caption{Visualization of latent variable activations.}
    \label{fig:visualization}
\end{figure}

To validate the amount of information disentanglement, we analyze the statistical properties of the learned latent variables $\bm{h}$, and whether the latent variables $\bm{h}$ learn spatial information independently from music content. 
In Fig. \ref{fig:a}, we show the activation of the mean of the latent distribution from two different panning configurations of the same song. The two uncorrelated graphs indicate that the $J=50$ dimensions learned in $\bm{h}$ differ from each other although the two versions share the exact same stem tracks. On the contrary, Fig. \ref{fig:b} shows highly correlated activation of two completely different songs, because their corresponding latent variables succeed to learn the same representation based on their same panning locations. These results suggest that the learned latent variables are influenced more by the spatial arrangement of the stem tracks and not by music content, achieving the desired disentanglement. 

Another visualization of the latent variables further strengthens our claim. We prepare a set of data samples, which are created from the combination of five different songs, and five different spatial configurations. If the latent space reflects the spatial information more than the music content, there must be a latent structure that reflects the spatial characteristics more. In Fig. \ref{fig:c} and \ref{fig:d}, we show a dimension-reduced latent space via principal component analysis. In \ref{fig:c}, the coloring is based on the songs, while in \ref{fig:d} it is based on the spatial configurations. We can see that the segments from the same spatial configuration form a cluster in \ref{fig:d}, while segments that belong to the same song are scattered everywhere. The average standard deviation for each color group in \ref{fig:c} is relatively higher than the ones in \ref{fig:d}, i.e., $0.060 > 0.038$. This observation again indicates that the encoder extracts music-invariant spatial features successfully.

\subsection{Style Transfer-Based Upmixing Results}

We conduct the style transfer-based upmixing and compare the model's performance against the baseline model. To this end, we prepare 540 test examples that are 10 seconds long and are rendered in one of the 20 arbitrarily defined panning configurations. 


\begin{table}[t]
\centering
\caption{Objective performance of the style transfer method.}
 
\resizebox{0.9\columnwidth}{!}{
\begin{tabular}{cccc}
  \noalign{\hrule height 1pt}
    & SD-SDR (dB) & $\Delta$ Angle ($^\circ$) & WILD\\
    \hline
    Style-transfer & \textbf{8.71$\pm$1.50} & \textbf{19.5$\pm$15} & \textbf{58.02$\pm$13.78}  \\
    \hline
    Baseline & 4.20$\pm$0.71 & 49.75$\pm$19.75 & 70.84$\pm$10.03 \\ 
  \noalign{\hrule height 1pt}
 \end{tabular}}
 \label{tab:data}
  \vspace{-0mm}
\end{table}

\begin{figure}[t]
     \centering
     \includegraphics[width=0.8\columnwidth]{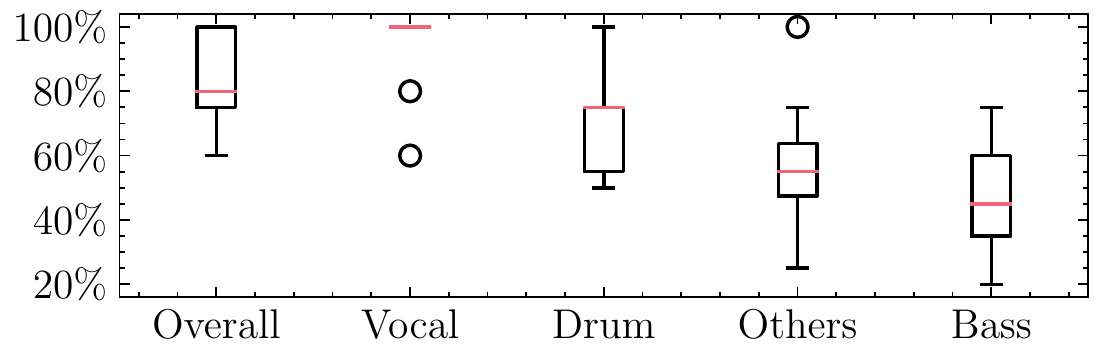}
     \vspace{-0.1in}
     \caption{Subjective tests on the rendered multichannel samples. }
     \label{fig:boxplot}
\end{figure}

\begin{itemize}[noitemsep,topsep=0pt, leftmargin=0in, itemindent=.15in]
    \item \textbf{Objective evaluations}: Table \ref{tab:data} summarizes the objective evaluation results. Our proposed style transfer-based upmixing method shows an obvious advantage over the baseline upmixing model with a better reconstruction score, less difference in angle, and WILD. It is noteworthy that the style-transferred results bear higher SD-SDR than the fixed-panning-based baseline, even in the presence of the machine learning model's algorithmic artifact. The SD-SDR score indicates that our model can properly fulfill the mission of arbitrarily positioning instruments during 5-channel generation without significant harm to the sound quality.
    
    \item \textbf{Subjective evaluations}: Fig. \ref{fig:boxplot} reports the percentage of subjects preferring the proposed style transfer-based upmixing over the baseline, regarding the direction similarity. Ten listening experts participated in our subjective test. More than 80\% of participants believe the proposed model is more similar to the reference, in terms of the overall perceptual panning. When it comes to the individual instruments, the reconstruction of \texttt{vocals} is the most favored, which might account for the overall performance due to its dominance in pop music. Our model gains preference from the listeners on \texttt{drums} and \texttt{other}, although with higher variance on the latter one, while the advantage doesn't pass to \texttt{bass}. We believe that the different preference on different instruments results from the fact that the decoder implicitly performs instrument separation and relocation for estimated sources: when an instrument is difficult to separate from the mixture, its relative scale in each channel and the resulting perceived direction in the 5-channel may be negatively affected, which also explains the phenomenon that our preference scores align to the diverse instrument-specific separation performances on the MUSDB18 dataset in the literature \cite{StollerD2018waveunet}. 



\end{itemize}

\subsection{Blind upmixing and other experiments}

To test our model's potential as a generator, we apply blind upmixing to MUSDB18 signals and to real-world music signals from our internal dataset. Because all the models are initially trained on MUSDB18, to let the model better accommodate real-world music signals, we fine-tune the model with real-world 5-channel music before blindly generating 5-channel examples. Due to the fact that the blind upmixing process is absolutely random, we cannot validate its quality and the surrounding effect via a quantitative measure. However, we find that the upmixing results are convincing based on informal listening by our peers. We provide some blind upmixing samples, both from MUSDB18 and the real-world dataset at \url{https://saige.sice.indiana.edu/research-projects/generative-upmixing}.

We evaluated different latent dimensions ranging from $J=20$ to $J=60$, while there is no strong enough trend to recommend a single number. We eventually chose $J=50$ for all the experiments.

\section{Conclusion}
In this paper, we proposed to tackle the stereo upmixing problem in a generative way, where spatial images and music content in the 5-channel signal are disentangled to allow style transfer-based upmixing. We formulated the problem into a modified VAE model and train the latent space to capture music-invariant spatial information, e.g., panning locations. Our experiments showed that the learned latent variables successfully capture spatial information separately from the music contents. Both the objective and subjective evaluations demonstrate that our style transfer-based upmixing model achieved interactive upmixing.

\bibliography{mjkim.bib}
\bibliographystyle{IEEEbib}

\end{document}